\documentclass{article}

\usepackage{arxiv}

\usepackage[utf8]{inputenc} 
\usepackage[T1]{fontenc}    
\usepackage{hyperref}       
\usepackage{url}            
\usepackage{booktabs}       
\usepackage{amsfonts}       
\usepackage{nicefrac}       
\usepackage{microtype}      
\usepackage{lipsum}
\usepackage{amsmath}
\usepackage{multirow}
\usepackage[pdftex]{graphicx}

\title{Evaluation of vulnerability reproducibility\\ in container-based Cyber Range}

\author{
  Ryotaro Nakata\\
  Institute of Information Security\\

  Yokohama, Kanagawa, Japan \\
  \texttt{dgs184101@iisec.ac.jp} \\
   \And
 Akira Otsuka \\
Institute of Information Security\\

  Yokohama, Kanagawa, Japan \\
  \texttt{otsuka@iisec.ac.jp} \\
}

\begin{document}
\maketitle

\begin{abstract}
The cyber range is a practical and highly educational information security exercise system, but it has not been widely used due to its high introduction and maintenance costs.
Therefore, there is a need for a cyber range that can be adopted and maintained at a low cost.
Recently, container type virtualization is gaining attention as it can create a high-speed and high-density exercise environment. However, existing researches have not clearly shown the advantages of container virtualization for building exercise environments.
Moreover, it is not clear whether the sufficient vulnerabilities are reproducible, required to conduct incident scenarios in the cyber range.
In this paper, we compare container virtualization with existing virtualization type and confirm that the amount of memory, CPU, and storage consumption can be reduced to less than 1/10 of the conventional virtualization methods.
We also compare and verify the reproducibility of the vulnerabilities used in common exercise scenarios and confirm that 99.3\% of the vulnerabilities are reproducible.
The container-based cyber range can be used as a new standard to replace existing methods.
\end{abstract}

\keywords{Information security education \and cyber range \and container type virtualization \and Docker \and vulnerability}

\section{Introduction}
With the development of ICT technology, the scale and impact of cyber-attacks continue to increase worldwide.
The threat of cyber-attacks is increasing in Japan due to the incidents that directly impact social activities and daily life, such as the leakage of virtual currencies and fraudulent use of mobile payments.
On the other hand, the shortage of human resources for security is pointed out, and the government and higher education institutions are making various efforts for human resource development. Still, the quantitative and qualitative shortage has not been solved.\cite{cyexec}.

Information security education includes lectures, individual learning with tools to reproduce and experience vulnerabilities, and practical exercises with cyber ranges\cite{CyTrONE}.
The cyber range is a large scale exercise system for learning by experiencing real security incidents in an organization in a virtual environment that simulates a real-world system.
Although the educational effect is high, the cyber range's introduction and maintenance are millions of dollars.

Although the effectiveness and necessity of cyber range are recognized, it is not easy to introduce and maintain cyber range on its own. Some universities have adopted it with companies' help, but the cyber-range has only spread to a few.
As a result, a low-cost exercise environment is required, and efforts to use container type virtualization for building the environment have been attracting attention.

For example, Labtainers at the US Naval Academy provides an exercise environment created using containers, which can be deployed and studied on students' terminals. The platform is open to the public and allows us to develop and share new exercises\cite{labtainers}.
CyExec at AIIT in Japan has also paid particular attention to the high portability and ease of sharing containers and has been trying to expand the educational curriculum and practice contents through joint development and use while also considering safety and ethical aspects\cite{cyexec}.

However, these studies do not show the advantages of container type virtualization in concrete terms, and the scope of its use in cyber range environments is not clear, which is a barrier to developing and using the container-based cyber range.
As a result, most existing cyber-range products and educational research use other virtualization types\cite{CyTrONE}.

This paper identified the advantages of container virtualization over other virtualization types in cyber range exercises.
Furthermore, we report the results of a comprehensive comparison with other virtualization types through experimentation using vulnerability assessment tools and exploit modules to investigate whether vulnerabilities and incidents in common exercise scenarios can be accurately replicated in Cyber range exercises.

\section{The cyber range}
\label{sec:cyberrange}

\subsection{Definition of cyber range}
TThe cyber range is a system to build and provide an environment for information security exercises. For smooth implementation of the exercises, with the following functions.
\begin{itemize} 
\item Construction of a highly realistic system environment \par
The system environment used in the real world, such as PCs, servers, networks, and security devices, can be recreated.
\item Duplicate or replace the environment \par
Flexible and quick resetting, duplication, and replacement of the exercise environment depending on the exercise scenario and student.
\item Reproduction of vulnerabilities and incidents \par
Vulnerable environments and actual malware can be used to recreate security incidents and execute attack and defense scenarios.
\end{itemize}

These are realized in an environment built with virtualization technology\cite{2001.06681}.

\subsection{Typical cyber range scenarios}

\noindent In the cyber range exercise, various scenarios will be developed depending on the learning objectives and the student's level. Typical attack scenarios include the following.

\begin{enumerate}
    \item Vulnerability search for web-server by attackers
    \item Installing backdoors through attacks on identified vulnerabilities
    \item Network penetration through a backdoor
    \item Port scanning to other devices on the network
    \item Intrusion and privilege escalation to discovered vulnerable devices
    \item Obtaining DB password information
    \item Unauthorized access to the DB and acquisition of confidential information
\end{enumerate}

\begin{figure}
\centering
\includegraphics[width=15cm]{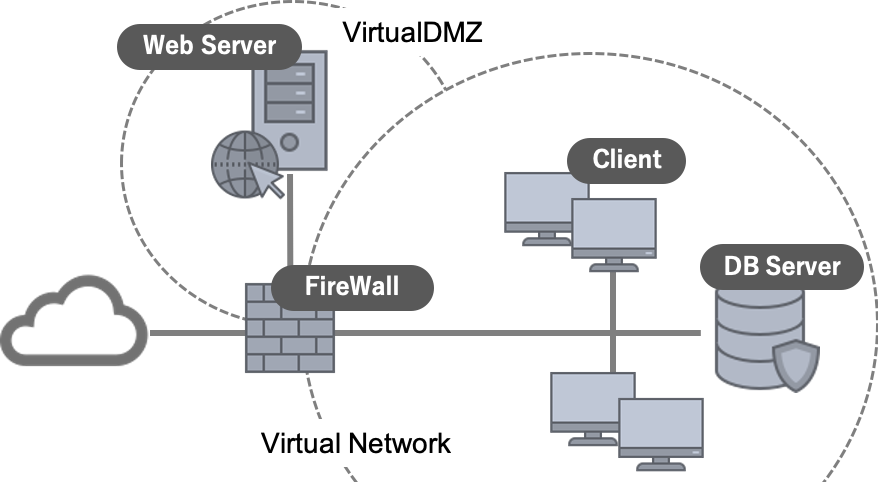}
\caption{Example of virtual network for cyber range}
\label{fig:virtualnet}
\end{figure}

Figure \ref{fig:virtualnet} shows an example of a network that assumes such an attack scenario.

In the cyber range, reproduce vulnerabilities in web servers, DB servers, client devices, etc. to carry out the scenario.
Also, security devices such as firewalls, IDS/IPS, etc. will be installed, depending on the exercise's nature.
To execute scenarios in the cyber range that replicate real security incidents and responses, prepare a virtual environment equivalent to the real system environment\cite{Needs}.

\section{Virtualization technology}

\subsection{Types of Virtualization Technologies}

\begin{figure}[h]
\centering
\includegraphics[width=10cm]{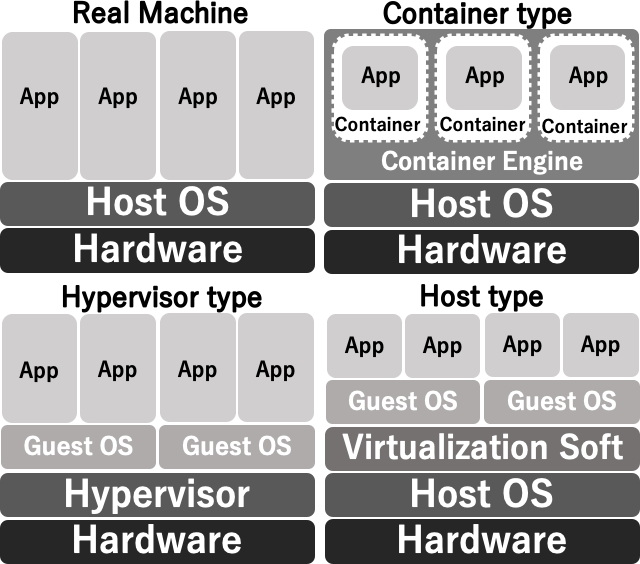}
\caption{Example of virtual network for cyber range}
\label{fig:virtual}
\end{figure}

Virtualization technology efficiently utilizes hardware by sharing and dividing the resources required for operating systems and applications among multiple environments (VM: Virtual Machine)\cite{1304.3557}. Figure \ref{fig:virtual} shows an overview of each virtualization technology types.

In the hypervisor type, a program called hypervisor builds a VM on specially prepared hardware, and it operates by occupying resources such as memory and CPU.  

In the host type, a VM is built by running dedicated software that plays a hypervisor role on an operating system (host OS) that is already running on the actual machine. A portion of the resources recognized by the host OS is allocated and operated.

Unlike other virtualization types, the container type operates like a VM by creating a separate namespace, called a container, on a running host OS that operated only the processes required for the functions to be used. Containers do not occupy physical resources and run as a single process on the host OS. Table \ref{tab:virtual} shows the characteristics of each virtualization types.

\begin{table}[th] 
\caption{Characteristics of each virtualization types}
\label{tab:virtual}
\centering
\begin{tabular}{c|ccc}\hline\hline
Virtualization & isolation & over & guest \\
type& level & head & OS\\\hline
Hypervisor & max & high & require \\
Host & high & max & require \\
Container & low & low & unrequire \\\hline
\end{tabular}
\end{table}

The isolation level indicates independence from the host OS and other VMs running on the same hardware. If the isolation level is low, there is a high probability that the host OS and other VMs will be affected if processing on the VM is slow or troubles occur. Overhead refers to the decrease in processing performance that occurs in a virtual environment. Since the hardware is accessed through the mechanism used to run the VM, there is a high probability that processing performance will decrease compared to the actual environment.

\begin{figure}[h]
\includegraphics[width=\linewidth]{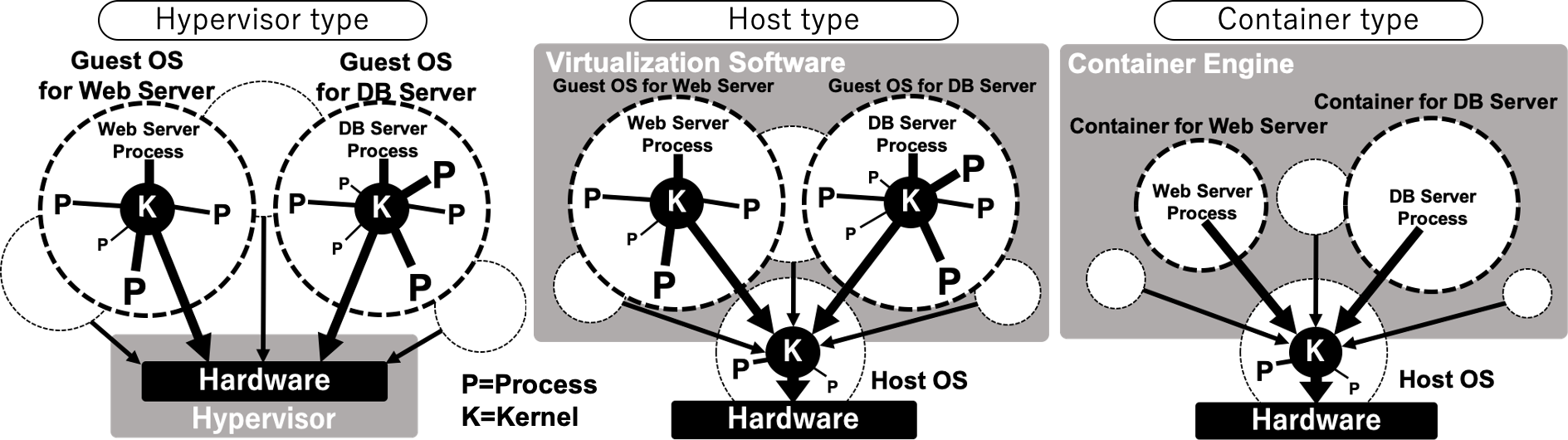}
\caption{Operating architecture of each virtualization types}
\label{fig:runimage}
\end{figure}

Figure \ref{fig:runimage} shows the operating architecture of each virtualization types.
In the hypervisor and host type, the VMs run independently and replicate the equivalent environment as the real machine. Still, they require the installation of a guest OS, which consumes a large amount of physical resources\cite{7921010}.
There is no require for the guest OS to run or to start and stop in the container type, and only the necessary functions can be run with minimal configuration. Containerized systems share kernels and resources. They can be fast and efficient. Still, they are affected by interactions with the host OS and other containers and may behave differently from the the real machine.\cite{7432984}.

\subsection{Virtualization type used in the cyber range}
 Existing cyber ranges use HV and hosted virtualization types.
Because the environment is nearly identical to that of a real machine, it can faithfully replicate the vulnerabilities and incidents.
In particular, virtualization solutions such as VMWare and Citrix are being used in the commercial cyber range for their stability and VM management capabilities\cite{vmware}.

However, depending on the number of participants and the number of exercise groups, there are more than 100 virtual instances running at the same time, requiring high load operations such as environment replication and rapid startup/termination. This is one of the reasons why the cost is so high because of the need for high-performance hardware\cite{cyexec}.

\begin{figure}[h]
\includegraphics[width=\linewidth]{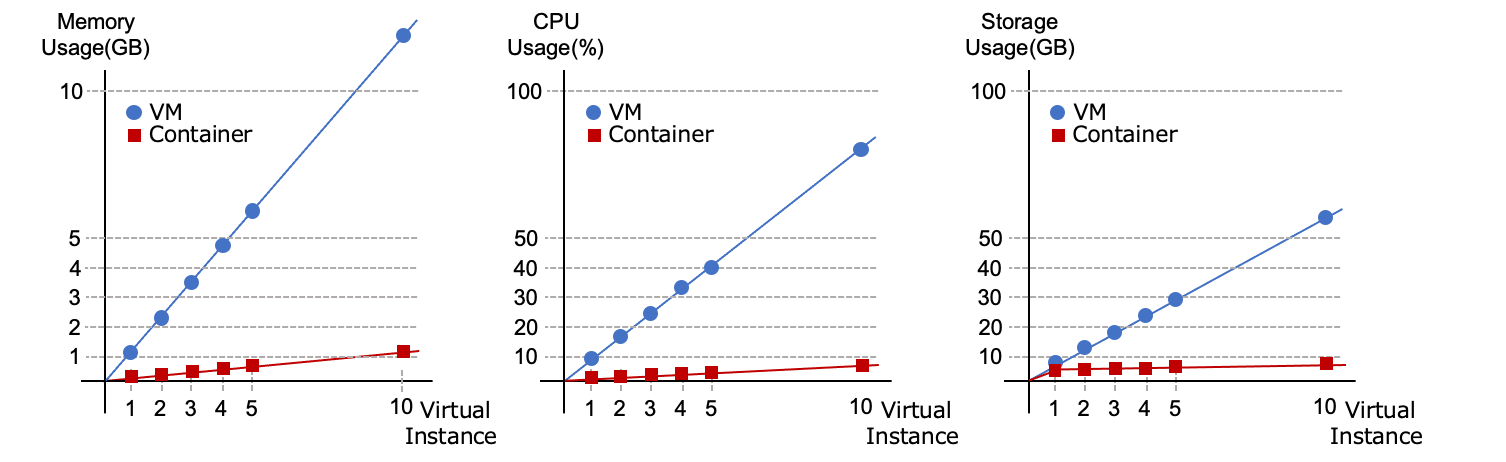}
\caption{Comparison of resource consumption as the number of virtual instances increases}
\label{fig:resource}
\end{figure}

\subsection{Advantages of Container-base cyber range}

\noindent cyber range environments using containerized virtualization are faster and less resource-intensive than other virtualization types.
Figure \ref{fig:resource} shows a comparison of the resources consumed by VMs and containers as the number of virtual instances increases.

In the example in the figure \ref{fig:resource}, the OS (Ubnutu16.04) was installed on a VM with 1GB memory allocated, and prepared as device that was intended to be a client for the cyber range environment.
We also installed the same OS and desktop package as the VM in a container running on the same host.
The resource consumption was compared by preparing the container as a container that allowed the same operations as the VM.

VMs always consume host resources without any particular actions, such as running a guest OS and various services.
Also, VMs consume more resources than the number of virtual instances because they allocate available resources in advance.
Many processes that are not necessary to execute cyber range scenarios are also running, and even virtual machines that are not specifically running consume a certain amount of resources, which is inefficient.

Containers, on the other hand, consume very few resources per instance because they run on a minimal number of processes.
Therefore, even if the cyber range environment is built and the number of virtual instances increases, the physical resource consumption can be significantly reduced compared to a VM environment. 
By using containers to build a cyber range environment, the required specifications of the host machine can be significantly reduced, and the environment can be built at a low cost.

\subsection{Concerns of Container}
\noindent Using container type virtualization is able to build a cyber range environment at a lower cost. However, container type virtualization has different characteristics than VMs used in the existing cyber range and may have different states and behaviors in executing scenarios.
As a result, individual vulnerabilities and incidents cannot be reproduced correctly and may not work as envisioned. The biggest concern with the container-based cyber range is whether the vulnerabilities required to execute a scenario can be reproduced on a container similar to on a VM.

\section{Experiments on Vulnerability Reproducibility}
\subsection{Reproducibilty Metrics}
\noindent To assess the reproducibility of vulnerabilities on a container-based cyber range, we model the OS/HW environment in which the programs as an oracle $\mathcal{O}$ and every system calls invoked by a program $A$ is sent to the oracle $\mathcal{O}$. 

$A^{\mathcal{O}_{\text {Real}}}$, $A^{\mathcal{O}_{\text{VM}}}$ and $A^{\mathcal{O}_{\text {Container }}}$

If a program is run in the environment over Real, VM, container, we write the output of the programs as $A^{\mathcal{O}_{\text {Real}}}$, $A^{\mathcal{O}_{\text{VM}}}$ and $A^{\mathcal{O}_{\text {Container }}}$ respectively. Where “Real" means physical environment where no virtualization technology is used. 

If the execution results of all programs are the same, there is no problem in executing any exercise scenario and it eliminates concerns of container-type cyber range.
Theoretically, the identification algorithm $\phi$ can be defined as the inability to identify in which environment the program $A$ was executed.

We write the set of results of running the program in each environment as $A^{Real}$, $A^{VM}$, and $A^{Container}$, respectively, as shown in the following equation.

\begin{displaymath}
\label{Equ:real}
A^{Real}=\left\{A \mid \phi\left(A^{O_\text{{Real}}}\right)=1\right\} 
\end{displaymath}
\begin{displaymath}
\label{Equ:real}
A^{VM}=\left\{A \mid \phi\left(A^{O_\text{{VM}}}\right)=1\right\}
\end{displaymath}
\begin{displaymath}
\label{Equ:real}
A^{Container}=\left\{A \mid \phi\left(A^{O_\text{{Container}}}\right)=1\right\}
\end{displaymath}

\begin{figure}[h]
\centering
\includegraphics[width=10cm]{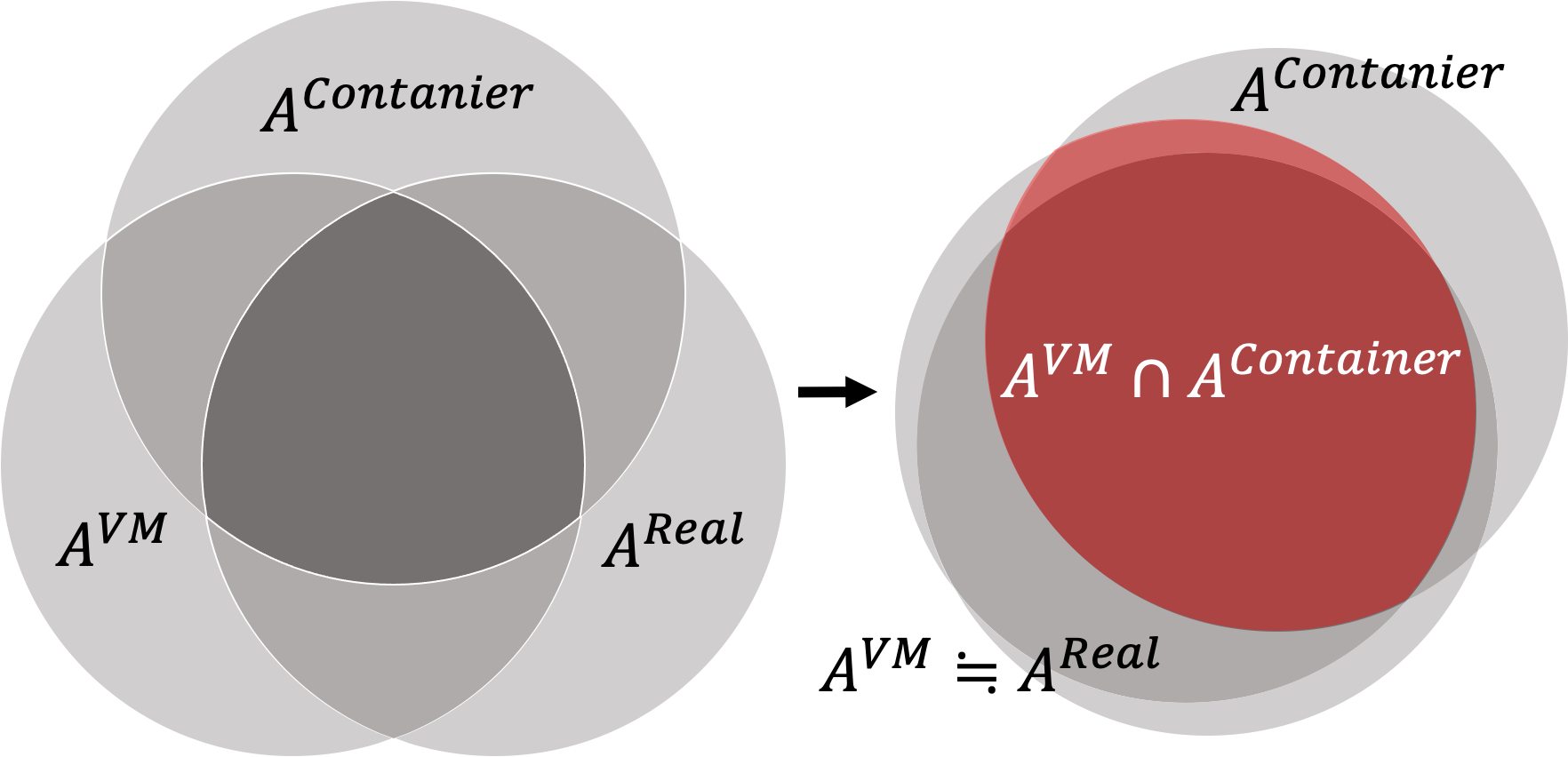}
\caption{Reproducibility Comparison Model}
\label{fig:model}
\end{figure}

By measuring the similarity of these sets, we confirm the reproducibility of vulnerabilities in the container-based cyber range.
The relationship between each set is shown in Figure \ref{fig:model}

Considering the existing use of VMs in the cyber range, we will consider Real and VM's to be nearly identical.
Thus, $J\left(A^{VM}, A^{\text {container}}\right)=\frac{\mid A^{VM} \cap A^{\text {Container}}}{\left|A^{VM} \cup A^{\text {Container}}\right|}$, which represents the similarity between VMs and containers, be the metric for using container type virtualization in the cyber range.

The higher the value of J, the more containers can be used in the same way as VMs in many scenarios, which eliminates concerns about container-based cyber range.

In reality, there are programs whose execution results are container-specific, such as programs related to physical vulnerabilities. Such programs should be excluded from the container-based cyber range and will be discussed in detail in Chapter 6.

\subsection{Experimental Method}

\noindent We will compare container and VM environment through an exhaustive experiment with programs that may be used in cyber range exercise scenarios.
We build an equivalent environment with VMs and containers to see if there is a difference between the scan results by the vulnerability assessment tool and the results of attacks against the detected vulnerabilities. 
Table \ref{tab:tool} shows the vulnerability assessment tools used in the experiment.

\begin{table}[th]
\caption{Vulnerability inspection tool used for verification}
\label{tab:tool}
\centering
\begin{tabular}{c|cccc}
\hline \hline
\multirow{2}{*}{Name} & \multicolumn{4}{c}{Target} \\ \cline{2-5} 
 &\begin{tabular}{c}Web\\App\end{tabular}&\begin{tabular}{c}Middle\\ware\end{tabular}&OS&\begin{tabular}{c}Net\\work\end{tabular}\\ \hline
OpenVAS &  & $\circ$ & $\circ$ & $\circ$ \\ \hline
Nmap &  &  &  & $\circ$ \\ \hline
Owasp ZAP & $\circ$ &  &  &  \\ \hline
Nikto2 & $\circ$ & $\circ$ &  &  \\ \hline
\end{tabular}
\end{table}

We have selected tools that can be used free in cyber range exercises.
Also, we divided the target areas into four areas: web applications, middleware, OS, and networks, and experimented with each tool. 
Using tools capable of scanning for each area, we thought we could perform a comprehensive experiment for various vulnerabilities.

\subsection{Experimental Settings}
\noindent We verified this with a Docker container and a VirtualBox VM.
Docker is a platform for container type virtualization, which is becoming more popular for various applications such as cloud services.
VirtualBox is a free hosted virtualization software, which is widely used in verification environments, education and research because of its ease of installation and the large number of supported operating systems.
We determined that the environments available for verification and defense (detection) of attacks are readily available and can be effectively verified for each.
The environments used for verification are shown in Table \ref{tab:env}.

\begin{table}[th]
\centering
\caption{Experimental Settings detail}
\label{tab:env}
\begin{tabular}{c|lc}
\hline\hline
env & \multicolumn{1}{c}{\begin{tabular}{c}Hardware\\ VM software and Image\end{tabular}} & Spec \\ \hline
Host & \begin{tabular}{l}\textbf{MacBookPro-13inch}\\ Ubuntu 18.04 LTS\end{tabular} & \begin{tabular}{c}2.3GHz Corei5\\ 16GB RAM\end{tabular} \\ \hline
VM& \begin{tabular}{l}\textbf{VirtualBox6.0}\\ Metasploitable2\\Metasploitable3\\ Kalilinux2019.1\end{tabular} & \begin{tabular}{c}1CPU\\ 2GB RAM\end{tabular} \\ \hline
{\begin{tabular}{c}cant-\\ainer\end{tabular}} & \multicolumn{2}{l}{\begin{tabular}{l}\textbf{DockerCE18.09}\\ Metasploitable2(created from VM image)\\ Metasploitable3(created from VM image)\\ kalilinux/kali-linux-docker\end{tabular}} \\ \hline
\end{tabular}
\end{table}

We used Kali-linux, a Linux distribution for penetration testing, and Metasploitable2 and Metasploitable3, which are used as a verification environment for deliberately adding vulnerabilities to containers and VMs, and compared the results of vulnerability assessment and exploit tests\cite{kali,metasploitable2}.

Since official container images of Metasploitable2 and Metasploitable3 are not available to the public, I created container images from each VM's configuration file.
By using the docker import command, which collects all files except /boot, /dev, /mnt, /proc, /sys, and /tmp, which are unnecessary for the operation of the containers, and creates a base container image from the archive files, I was able to create my own container image. We have created an environment that works with the same configuration, although the startup process is different.

By using the docker import command, which collects all files using tar command except /boot, /dev, /mnt, /proc, /sys, and /tmp, which are unnecessary for the operation of the containers, and creates a base container image from the archive files, I was able to create my own container image. We have created an environment that works with the same configuration, although the startup process is different.

\begin{figure}[h]
\centering
\includegraphics[width=10cm]{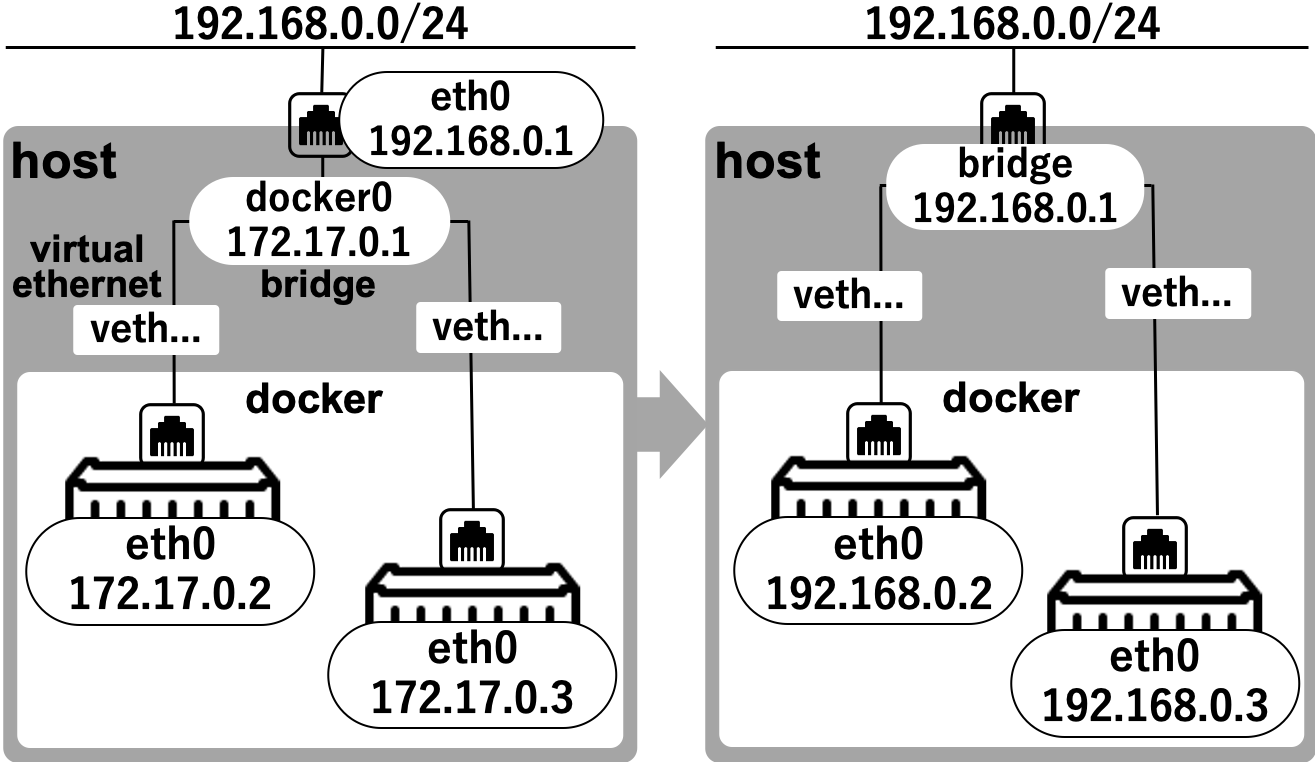}
\caption{Network Settings on Docker}
\label{fig:network}
\end{figure}

The network in the verification environment does not use the standard Docker network. As shown in the figure \ref{fig:network}, containers were placed on the same segment as the host, and no communication restrictions were placed between the host and containers.
In the cyber range exercise, it is necessary to construct a flexible network environment to reproduce the real environment.

\section{Experimantal Results}
\subsection{Measuring Reproducibility with Vulnerability Assesment Tools}

\subsubsection{OpenVAS}

\noindent OpenVAS is an open-source vulnerability testing tool with rich testing capabilities. It can detect a wide range of vulnerabilities such as software bugs, usage flaws, and configurations, and can check for corresponding CVEs (Common Vulnerabilities and Exposures)\cite{cve} using the latest database\cite{openvas}.
Table \ref{tab:openvas} compares the results of OpenVAS CVE-based vulnerability checks by the number of detections per VM and container environment. Since the number of CVE detections is very high, it is aggregated by CWE (Common Weakness Enumeration) \cite{cwe}.

\begin{table}
\centering
\caption{Comparison of the number of vulnerabilities detected by OpenVAS}
\label{tab:openvas}
\renewcommand{\arraystretch}{0.95}
\begin{tabular}{clrr}
\hline\hline
CWE & \multicolumn{1}{c}{\begin{tabular}{c}vulnerability classification\end{tabular}} & VM & \begin{tabular}[c]{@{}l@{}}cont-\\ ainer\end{tabular}  \\ \hline
16 & Configuration & 2 & 2 \\ \hline
17 & Code &2 & 2 \\ \hline
18 & Source Code &1 &1 \\ \hline
20 & Improper Input Validation & 82 & 82 \\ \hline
22 & Pass Traversal & 7 & 7 \\ \hline
59 & Link Following & 6 &6 \\ \hline
74 & Injection &1 &1 \\ \hline
79 & Cross Site Scripting& 33 & 33 \\ \hline
89 & SQL Injection & 7& 7\\ \hline
93 & CRLF Injection & 4& 4 \\ \hline
94 & Code Injection & 13 &13 \\ \hline
113 & HTTP Response Splitting & 1 & 1 \\ \hline
119 & Buffer Error & 66& 66 \\ \hline
125 & Out-of-bounds Read & 2& 2 \\ \hline
134 &  \begin{tabular}[c]{@{}l@{}}Use of Externally-Controlled\\  Format String\end{tabular}  & 5 & 5 \\ \hline
189 & Numeric Errors & 31 & 31 \\ \hline
190 & Integer Overflow or Wraparound& 3& 3 \\ \hline
200 & \begin{tabular}[c]{@{}l@{}}Exposure of Sensitive Information\\
 to an Unauthorized Actor\end{tabular} & 42& 42 \\ \hline
254 & 7PK - Security Features & 2&2\\ \hline
255 & Credentials Management Errors & 2& 2 \\ \hline
264 & \begin{tabular}[c]{@{}l@{}}Permissions, Privileges, \\
and Access Controls\end{tabular}  & 54& 54 \\ \hline
275 & Permission Issues & 1 & 1 \\ \hline
284 & Improper Access Control & 5 &5 \\ \hline
287 & Improper Authentication & 8 &8 \\ \hline
295 & Improper Certificate Validation &2 &2 \\ \hline
310 & Cryptographic Issues & 22& 22 \\ \hline
311 & \begin{tabular}[c]{@{}l@{}}Missing Encryption \\of Sensitive Data\end{tabular}   & 22& 22 \\ \hline
320 & Key Management Errors & 2& 2 \\ \hline
327 & \begin{tabular}[c]{@{}l@{}}Use of a Broken or Risky \\ Crypto-graphic Algorithm\end{tabular} & 1 & 1 \\ \hline
345 & \begin{tabular}[c]{@{}l@{}}Insufficient Verification \\ of Data Authenticity\end{tabular} & 1 & 1 \\ \hline
352 & Cross Site Request Forgery & 5& 5 \\ \hline
362 & Race Condition' & 9 & 9 \\ \hline
384 & Session Fixation & 1& 1\\ \hline
399 & Resource Management Errors & 51& 51 \\ \hline
400 & \begin{tabular}[l]{@{}l@{}}Uncontrolled Resource\\   Consumption\end{tabular}& 2&2\\ \hline
415 & Double Free & 1& 1\\ \hline
416 & Use After Free & 2 & 2 \\ \hline
476 & NULL Pointer Dereference & 8& 8\\ \hline
502 & \begin{tabular}[l]{@{}l@{}}Deserialization of Untrusted\\ Data\end{tabular} & 2& 2\\ \hline
552 & \begin{tabular}[c]{@{}l@{}}Files or Directories Accessible \\ to External Parties\end{tabular} & 1 & 1 \\ \hline
601 & Open Redirect & 5 &5 \\ \hline
732 & \begin{tabular}[c]{@{}l@{}}Incorrect Permission Assignment\\ for Critical Resource\end{tabular} & 1 & 1 \\ \hline
772 & \begin{tabular}[c]{@{}l@{}}Missing Release of Resource\\  after Effective Lifetime\end{tabular} & 1 & 1 \\ \hline
787 & Out-of-bounds Write & 1 & 1 \\ \hline
835 & Infinite Loop & 8& 8 \\ \hline
Design& Design errors & 4 & 4 \\ \hline
Other & Other errors & 50 & 50 \\ \hline
noinfo & Lack of information & 123 & 123 \\ \hline
\end{tabular}
\end{table}

Initially, several CVE items were not detected by the container. However, by checking the contents of the vulnerability, starting the services that were not working on the container side, and checking and correcting the settings, we were able to achieve complete agreement on all items. Therefore, we were able to confirm the same reproducibility in VM and container environments for the items confirmed by the vulnerability test using OpenVAS.

\subsubsection{Nmap}

Nmap is an open-source port scanner with extensive OS and service version detection capabilities\cite{nmap}.
On the other hand, it is also used, for example, by malicious people to investigate the status of a host that is under attack\cite{nmapattack}.
In cyber-range exercises, they are often used in the early stages of exercise scenarios, for example, to respond to incidents after port scan detection.
Therefore, experimental results from Nmap are also important in a container-based Cyber-range environment. Table \ref{tab:nmap} shows the comparative results of assessment by NMAP.

\begin{table}
\centering
\caption{Comparison of detection results by Nmap}
\label{tab:nmap}
\begin{tabular}{cllcc}
\hline\hline
 \begin{tabular}[c]{@{}c@{}}CWE\\ex.\end{tabular}&port & version & VM & \begin{tabular}[c]{@{}l@{}}cont-\\ainer\end{tabular} \\ \hline
\begin{tabular}[c]{@{}l@{}}189\\339\end{tabular}&\begin{tabular}[c]{@{}l@{}}21/\\ftp\end{tabular} & \begin{tabular}[c]{@{}l@{}}vsftpd 2.3.4\\ProFTPD 1.3.5\end{tabular}& \hfil $  \circ $ \hfil & \hfil $  \circ $ \hfil \\ \hline
\begin{tabular}[c]{@{}l@{}}119\\200\end{tabular}& \begin{tabular}[c]{@{}l@{}}22/\\ssh\end{tabular} & \begin{tabular}[c]{@{}l@{}}OpenSSH 4.7p1\\ OpenSSH 6.6 1p1\end{tabular} & \hfil $  \circ $ \hfil & \hfil $  \circ $ \hfil \\ \hline
\begin{tabular}[c]{@{}l@{}}254\\416\end{tabular}&\begin{tabular}[c]{@{}l@{}}53/\\dns\end{tabular}& ISC BIND 9.4.2  & \hfil $  \circ $ \hfil & \hfil $  \circ $ \hfil \\ \hline
\begin{tabular}[c]{@{}l@{}}79\\287\end{tabular} &\begin{tabular}[c]{@{}l@{}}80/\\http\end{tabular} & \begin{tabular}[c]{@{}l@{}}Apache 2.2.8\\Apache 2.4.7\end{tabular}   & \hfil $  \circ $ \hfil & \hfil $  \circ $ \hfil \\ \hline
399& \begin{tabular}[l]{@{}l@{}}111/\\rpcbind\end{tabular} & 2 (RPC \#100000) & \hfil $  \circ $ \hfil & \hfil $  \circ $ \hfil \\ \hline
\begin{tabular}[c]{@{}l@{}}22\\275\end{tabular} & \begin{tabular}[l]{@{}l@{}}139,\\445/\\samba\end{tabular} &  \begin{tabular}[c]{@{}l@{}}Samba smbd\\ 3.X-4.X\end{tabular} & \hfil $  \circ $ \hfil & \hfil $  \circ $ \hfil \\ \hline
\begin{tabular}[c]{@{}l@{}}264\\290\end{tabular} &\begin{tabular}[c]{@{}l@{}}631/\\ipp\end{tabular}& CUPS 1.7& \hfil $  \circ $ \hfil & \hfil $  \circ $ \hfil \\ \hline
\begin{tabular}[c]{@{}l@{}}94\\119\end{tabular} & \begin{tabular}[l]{@{}l@{}}1099/\\java\\-rmi\end{tabular} & Java Rmi Registry& \hfil $  \circ $ \hfil & \hfil $  \circ $ \hfil \\ \hline
264&\begin{tabular}[c]{@{}l@{}}2049/\\nfs\end{tabular}& 2-4(RPC \#100003)  & \hfil $  \circ $ \hfil & \hfil $  \circ $ \hfil \\ \hline
\begin{tabular}[c]{@{}l@{}}22\\399\end{tabular} &\begin{tabular}[c]{@{}l@{}}2121/\\ftp\end{tabular}& ProFTPD 1.3.1 & \hfil $  \circ $ \hfil & \hfil $  \circ $ \hfil \\ \hline
\begin{tabular}[c]{@{}l@{}}134\\189 \end{tabular}& \begin{tabular}[l]{@{}l@{}}3306/\\MySQL\end{tabular}& \begin{tabular}[c]{@{}l@{}}MySQL 5.0.51a\\MySQL 5.5.62\end{tabular} & \hfil $  \circ $ \hfil & \hfil $  \circ $ \hfil \\ \hline
20& \begin{tabular}[l]{@{}l@{}}3500,\\8181\\/http\end{tabular} & \begin{tabular}[l]{@{}l@{}}Webrick httpd\\/1.3.1\end{tabular}  & \hfil $  \circ $ \hfil & \hfil $  \circ $ \hfil \\ \hline
other& \begin{tabular}[l]{@{}l@{}}3632/\\distccd\end{tabular} & distccd v1 & \hfil $  \circ $ \hfil & \hfil $  \circ $ \hfil \\ \hline
\begin{tabular}[c]{@{}l@{}}264\\284\end{tabular} & \begin{tabular}[l]{@{}l@{}}5432/\\Post\\-greSQL\end{tabular}& \begin{tabular}[c]{@{}l@{}}PostgreSQL\\DB8.3.0-8.3.7\end{tabular} & \hfil $  \circ $ \hfil & \hfil $  \circ $ \hfil \\ \hline
other&\begin{tabular}[c]{@{}l@{}}5900/\\vnc\end{tabular}& VNC protocol 3.3  & \hfil $  \circ $ \hfil & \hfil $  \circ $ \hfil \\ \hline
\begin{tabular}[c]{@{}l@{}}20\\189\end{tabular} & \begin{tabular}[l]{@{}l@{}}6667,\\6697/\\irc\end{tabular} & Unreal ircd & \hfil $  \circ $ \hfil & \hfil $  \circ $ \hfil \\ \hline
16& \begin{tabular}[l]{@{}l@{}}8009/\\aip13\end{tabular}& \begin{tabular}[c]{@{}l@{}}Apache Jserv\\ (Protocol v1.3)\end{tabular} & \hfil $  \circ $ \hfil & \hfil $  \circ $ \hfil \\ \hline
\begin{tabular}[c]{@{}l@{}}20\\119\end{tabular} & \begin{tabular}[l]{@{}l@{}}8180/\\http\end{tabular}& \begin{tabular}[c]{@{}l@{}}Tomcat/Coyte\\ JSP engine 1.1\end{tabular} & \hfil $  \circ $ \hfil & \hfil $  \circ $ \hfil \\ \hline
\begin{tabular}[c]{@{}l@{}}189\\399\end{tabular} &\begin{tabular}[c]{@{}l@{}}8787/\\drb\end{tabular}& Ruby DRb RMI & \hfil $  \circ $ \hfil & \hfil $  \circ $ \hfil \\ \hline 

\end{tabular}
\end{table}

In the Nmap experiment, we performed an optional scan to detect the type and version of the operating system and running services, and compared the detection results.
As with OpenVAS, there were some services that could not be detected at first, especially those related to remote connections, but we were able to obtain the same detection results by starting the missing services and adding settings, and we were able to confirm the same reproducibility between VM and container environments. In addition, we checked the vulnerabilities that could exist in the detected contents and confirmed the corresponding CWE.

As a side note, the first scan with Nmap on the VM took 138.93 seconds, compared to 113.75 seconds in the container environment. We were able to perform several scans in about the same amount of time, and the container was able to perform the scans at a faster.

\subsubsection{OWASP ZAP}

ZAP (Zed Attack Proxy) is an open-source web application vulnerability testing tool provided by OWASP (Open Web Application Security Project) that, like Nmap, can detect many vulnerabilities and check how to deal with them.
Metasploitable2 and Metasploitable3, which are prepared as vulnerable environments, have several web application environments, and web pages with various vulnerabilities, can be checked. Table \ref{tab:zap} shows a comparison of the results of the ZAP vulnerability check\cite{7340766}. 
For Metasploitable2 and Metasploitable3, there was a significant difference in the number of vulnerabilities detected.
It is probably due to the different software and versions that make up the web application.

\begin{table}[t]
\centering
\caption{Comparison of the number of detected vulnerabilities by ZAP}
\label{tab:zap}
\begin{tabular}{clrrrr}
\hline\hline
\multirow{3}{*}{CWE}&\multicolumn{1}{c}{\multirow{3}{*}{Vulnerability}} & \multicolumn{2}{c}{ \begin{tabular}[c]{@{}c@{}}Metasploi-\\table2\end{tabular}} & \multicolumn{2}{c}{\begin{tabular}[c]{@{}c@{}}Metasploi-\\table3\end{tabular}} \\ \cline{3-6} 
\multicolumn{1}{c}{} & & VM &\!\begin{tabular}[c]{@{}l@{}}cont-\\ainer\end{tabular} &VM & \begin{tabular}[c]{@{}l@{}}cont-\\ainer\end{tabular} \\ \hline
89&SQL Injection & 358 & 422 &2&2\\ \hline
97&Server Side Include & 1 & 1 &0&0\\ \hline
79&XSS(refrected) & 1075 & 1000 &1&1\\ \hline
79&XSS(stored) & 5 & 5 &0&0\\ \hline
22&Pass Traversal & 21 & 21 &0&0\\ \hline
78&Command Injection & 361 & 342 &0&0\\ \hline
98&File Inclusion & 209 & 206 &0&0\\ \hline
200& \begin{tabular}[c]{@{}l@{}}Application error\\  disclosure\end{tabular}& 242 & 246 &1&1\\ \hline
548&Directory Browsing & 14 & 15 &21&21\\ \hline
472&\begin{tabular}[c]{@{}l@{}}Parameter\\tampering \end{tabular} & 13 & 14 &1&0\\ \hline
200&\begin{tabular}[c]{@{}l@{}}Buffer Error \\disclosure \end{tabular}& 291 & 287 &1&1\\ \hline
200&Private IP disclosure & 136 & 139 &1&1\\ \hline
\end{tabular}
\end{table}

In the case of Metasploitable2, the overall trend is that the number of detections for each vulnerability is similar for VMs and containers. For cases where the number of detections by containers is lower than VMs, there may be factors that can be addressed to reproduce the same results, such as service start and configuration issues, or the presence or absence of required files. On the other hand, if the number of detections by containers is higher than the number of VMs, it is necessary to confirm whether the difference is due to an intrinsic difference caused by unreproducible vulnerabilities in the container environment resource-referencing behavior of the host OS. 

However, it was found that VMs and containers can reproduce the same vulnerabilities for many items. Therefore, the cyber range exercise can reproduce the targeted vulnerabilities and utilize them in the scenario.

\subsubsection{Nikto2}
Nikto2 is a dictionary-based tool for checking published security holes and problematic settings in web servers and web applications running on them. Table \ref{tab:nikto2} shows the comparative results of testing with Nikto2.

\begin{table}[h]
\centering
\caption{Comparison of vulnerability results by Nikto2}
\label{tab:nikto2}
\begin{tabular}{cclrr}
\hline\hline
CWE & OSVDB & Example of detection & VM & \begin{tabular}[c]{@{}l@{}}cont-\\ainer\end{tabular} \\ \hline
other&48 &directory is browsable & 1 & 1 \\ \hline
other&119 &CVE-1999-0269 & 2 & 2 \\ \hline
284&576 &\begin{tabular}[c]{@{}l@{}}Vulnerbility of \\weblogic\end{tabular}  & 1 & 1 \\ \hline
693&877 & vulnerbility of XST & 1 & 1 \\ \hline
other&3092 & \begin{tabular}[c]{@{}l@{}}phpMyAdmin is \\browsable\end{tabular}  & 7 & 7 \\ \hline
other&3233 & \begin{tabular}[c]{@{}l@{}}configuration file is \\browsable\end{tabular} & 3 & 3 \\ \hline
other&3268 & \begin{tabular}[c]{@{}l@{}}Directory index is \\browsable\end{tabular} & 9 & 9 \\ \hline
119&3288 &\begin{tabular}[c]{@{}l@{}}Vulnerbility of \\Abyss 1.03\end{tabular} & 1 & 1 \\ \hline
other&12184 &PHP version disclosure & 1 & 1 \\ \hline
\end{tabular}
\end{table}

Nikto2 can perform checks based OSVDB(Open Source Vulnerability DataBase). The OSVDB was operated as a database of vulnerability information as an impartial non-profit organization, but it has not been updated since its operation has been suspended. There was no problem in the operation of Nikto2 as a vulnerability testing tool. In this test, the same results were obtained for both VMs and containers, confirming the same reproducibility. Depending on the vulnerability of each OSVDB item, the corresponding CWE items were confirmed.

\subsection{Measuring Reproducibility with Exploit Modules}

\begin{table}[th]
\centering
\caption{Comparison of verification results by attack test}
\label{tab:metasploit}
\begin{tabular}{cllcc}
\hline\hline
CWE&\multicolumn{2}{c}{metasploit module} & \!V\!M & \begin{tabular}[c]{@{}c@{}}cont-\\ ainer\end{tabular}\\ \hline
\multirow{11}{*}{\begin{tabular}[c]{@{}c@{}}119\\ \multirow{4}{*}{200}\\\\\\\\ \multirow{2}{*}{310} \\ \\ \multirow{2}{*}{416}\\ \\ \multirow{3}{*}{other}\\ \\ \\ \end{tabular}} &\multirow{12}{*}{\begin{tabular}[c]{@{}l@{}}auxiliary\\ /scanner\end{tabular}} & ssl/openssl\_heartbleed & $  \circ $ & $  \circ $ \\ \cline{1-1}\cline{3-5} 
& & ssh/ssh\_enumusers & $  \circ $ & $  \circ $ \\ \cline{3-5} 
& & http/options & $  \circ $ & $  \circ $ \\ \cline{3-5} 
& & http/trace & $  \circ $ & $  \circ $ \\ \cline{3-5} 
& & http/tomcat\_enum & $  \circ $ & $  \circ $ \\ \cline{1-1}\cline{3-5} 
& & http/ssl\_version & $  \circ $ & $  \circ $ \\ \cline{3-5} 
& & ssl/openssl\_ccs & $  \circ $ & $  \circ $ \\ \cline{1-1}\cline{3-5} 
& & \begin{tabular}[c]{@{}l@{}}http/apache\\\_optionsbleed \end{tabular}& $  \circ $ & $  \circ $ \\ \cline{1-1}\cline{3-5} 
& & rservices/rexec\_login & $  \circ $ & $  \circ $ \\ \cline{3-5} 
& & rservices/rlogin\_login & $  \circ $ & $  \circ $ \\ \cline{3-5} 
& & rservices/rsh\_login & $  \circ $ & $  \circ $ \\ \hline
119&\multicolumn{2}{l}{\begin{tabular}[c]{@{}l@{}}auxiliary/server/openssl\\\_heartbeat\_client\_memory\end{tabular}}&$  \circ $&$  \circ $\\ \hline
\multirow{3}{*}{\begin{tabular}[c]{@{}c@{}}\multirow{2}{*}{94}\\\\\multirow{2}{*}{189}\end{tabular}} &\multirow{4}{*}{\begin{tabular}[c]{@{}l@{}}exploit\\ /linux\end{tabular}} &\multirow{2}{*}{\begin{tabular}[c]{@{}l@{}}samba\\/is\_known\_pipename\end{tabular} } &\multirow{2}{*}{$  \circ $} & \multirow{2}{*}{$  \circ $}  \\\\ \cline{1-1}\cline{3-5} 
& & \begin{tabular}[c]{@{}l@{}}samba\\/setinfopolicy\_heap\end{tabular} & $  \circ $ & $  \circ $ \\ \hline
\multirow{4}{*}{\begin{tabular}[c]{@{}c@{}}16\\\multirow{2}{*}{20}\\\\\multirow{2}{*}{284}\\\\other\end{tabular}}&\multirow{6}{*}{\begin{tabular}[c]{@{}l@{}}exploit\\ /unix\end{tabular}} & misc/distcc\_exec & $  \circ $ & $  \circ $ \\ \cline{1-1}\cline{3-5} 
& & \begin{tabular}[c]{@{}l@{}}irc/unreal\_ircd\_3281\\\_backdoor\end{tabular} & $  \circ $ & $  \circ $ \\\cline{1-1}\cline{3-5} 
& &\begin{tabular}[c]{@{}l@{}}ftp/proftpd\_modcopy\\ \_exec\end{tabular}  & $  \circ $ & $  \circ $ \\ \cline{1-1}\cline{3-5} 
& & webapp/twiki\_history & $  \circ $ & $  \circ $ \\ \hline
\multirow{4}{*}{\begin{tabular}[c]{@{}c@{}}\multirow{2}{*}{20}\\\\\multirow{2}{*}{other}\\\\noinfo\end{tabular}}&\multirow{5}{*}{\begin{tabular}[c]{@{}l@{}}exploit\\ /multi\end{tabular}} &\multirow{2}{*}{\begin{tabular}[c]{@{}l@{}}browser\\ /java\_storeimagearray\end{tabular}}  & \multirow{2}{*}{$  \circ $} & \multirow{2}{*}{$  \circ $} \\ 
\\\cline{1-1}\cline{3-5} & &\begin{tabular}[c]{@{}l@{}}http/php\_cgi\_arg\\ \_injection\end{tabular}  & $  \circ $ & $  \circ $ \\ \cline{1-1}\cline{3-5} 
& & samba/usermap\_script & $  \circ $ & $  \circ $ \\ \hline
\end{tabular}
\end{table}

\noindent The Metasploit framework included in Kali-linux includes a variety of exploit modules. By using them, we can conduct proof-of-concep and other experiments. We compared the results of the attack experiments against Metasploitable2 and Metasploitable3 in the VM and container environments to see if the attacks were successful and if the display and behavior were identical. Also, the exploit modules were selected from those capable of attacking vulnerabilities detected by the vulnerability assessment tool used in the 5.1 experiment.

Table \ref{tab:metasploit} shows the modules used in the attack experiments and the experiments' results.
All the exploit modules used worked as expected.
Also, the screen display and behavior were the same in both VM and container environments, and we were able to confirm the equivalent of reproducibility in the attack experiments.

\subsection{Overall Results}
\begin{figure}[h]
\includegraphics[width=\linewidth]{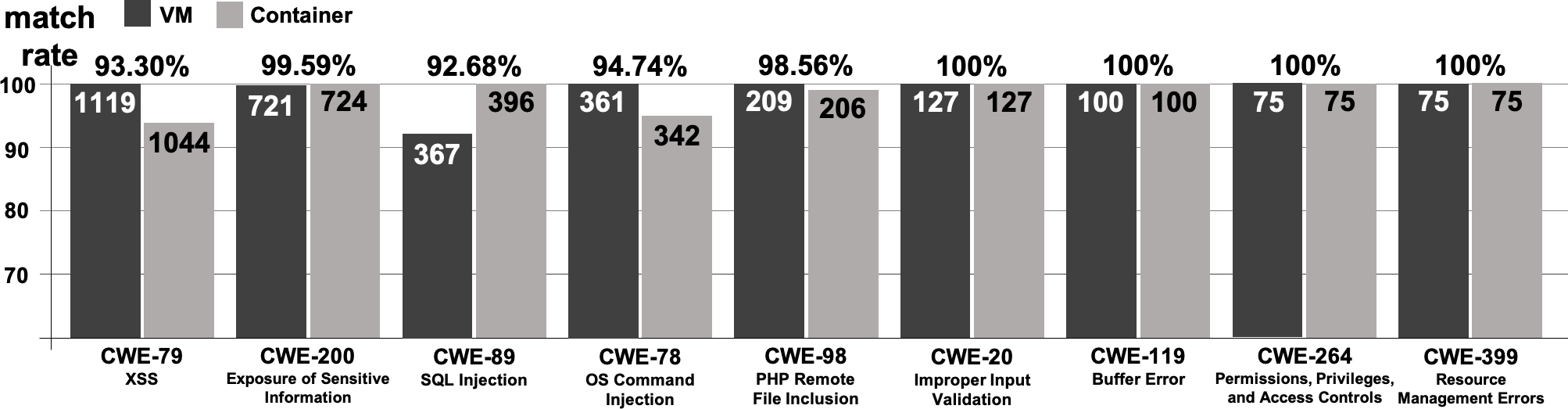}
\caption{Match rate of vulnerability reproducibility about VM and container}
\label{fig:matchrate}
\end{figure}

\noindent We checked the percentage of coincidence between VMs and containers for all experiments performed.
Figure \ref{fig:matchrate} shows the total number of detections for each CWE and the percentage of identical results between VMs and containers for the vulnerabilities and detection items identified in each experiment. Because the items are displayed in order of number, there are many CWE items that were detected but are not displayed, or that do not correspond to CWEs.

In this verification, the reproducibility of the ZAP vulnerability was not a 100\%  match, but the similarity $J$ calculated for all items was $0.993$. 

This value is higher than we expected, and we believe it is acceptable even if we assume the creation of a practical cyber range environment and the use of exercise scenarios. This validation largely eliminates the cyber range concerns of container-based virtualization.

In this experiment, initially, some of the results were not match between VMs and containers. However, most of them were related to starting and configuring services in a container environment. Therefore, if the files that make up the container are equivalent to those of a VM, the same reproducibility can be ensured by adding the appropriate configuration and startup process.
This experiment largely eliminates the cyber range concerns of container-based cyber range.


\section{Limitations of Reproducibility}
\noindent In container type virtualization, it may be identifiable as a container because the comparison of behavioral results with other virtualization schemes is inconsistent. For example, in the case of a direct attack on the part shared with the host OS. These are not suitable for the cyber range environment and should be excluded, but we checked if they are identifiable with the VM. 
Table \ref{tab:resource} shows the results of an attack on a VM and a container.

\begin{table}[h]
\centering
\caption{Physical resource consumption and behavior of virtual environment}
\label{tab:resource}
\begin{tabular}{ccl}
\hline\hline
resource & type & \multicolumn{1}{c}{results} \\ \hline
\multirow{4}{*}{memory} & VM & \begin{tabular}[c]{@{}l@{}}Program is delayed at VM \\memory allocation limit, but other \\VMs are not affected.\end{tabular} \\ \cline{2-3} 
 &\begin{tabular}[c]{@{}l@{}}cont-\\ ainer\end{tabular}& \begin{tabular}[c]{@{}l@{}}When physical memory usage \\reaches a limit, the entire operation, \\including the host OS, is delayed.\end{tabular} \\ \hline
\multirow{5}{*}{strage} & VM & \begin{tabular}[c]{@{}l@{}}Machine stops at VM storage\\ allocation limit, but other VMs are\\ not affected.\end{tabular} \\ \cline{2-3} 
 &\begin{tabular}[c]{@{}l@{}}cont-\\ ainer\end{tabular}& \begin{tabular}[c]{@{}l@{}}When physical storage usage \\reaches a limit, the entire operation, \\including the host OS, is stopped.\end{tabular} \\ \hline
\end{tabular}
\end{table}

A VM occupies a portion of the resources on the host OS and operates as an independent machine. Therefore, even if an attack slows down or stops the VM, it does not interfere with the host OS's operation. However, the standard configuration of the container shares resources with the host OS and other containers. An attack on the container can affect the entire system, including the host OS, which is a clear difference from the VM. 
Against such attacks, it was confirmed that container type virtualized environments are unsuitable for cyber range exercises. Similarly, vulnerabilities such as kernel vulnerabilities, host OS configuration, and files required for container execution, and attacks against them, should be excluded from container-based cyber range exercises.

However, Docker also has options for limiting shared resources, such as limited memory and storage consumption to a certain percentage of the host OS. If configured correctly, it may reduce the impact to an acceptable range on the cyber range.

\section{\uppercase{Conclusions}}

\noindent While there is a shortage of information security personnel, the cyber range is expected to be highly effective in education. However, due to cost issues and other factors, this system's introduction has been limited to a few cases.

Therefore, to disseminate an inexpensive and deployable cyber range using container-based virtualization, we confirmed the superiority of containers and the performance of reproducing vulnerabilities through comprehensive experiments.
Comparing the performance of containers and VMs in a typical cyber-range environment, the containers consumed less than 1/10th of the resources of the VMs. Containers can run more virtual instances than VMs on the same host, building a lower cost cyber range.

We also compared the reproducibility of vulnerabilities between containers and VMs in an exhaustive experiment using the vulnerability assessment tool and the exploit module, and found a very high similarity J of 0.997.
Although content derived from container characteristics must be excluded, containers have a very high vulnerability reproduction performance, confirming that container-based virtualization can be fully used in the cyber range.

These contents can be used as a benchmark for scenario development and exercise implementation for container-based cyber ranges such as CyExec. By the way, CyExec has already conducted the exercise.
In the future, we hope to promote the use of container-based cyber range through scenario development and exercises.
We will also promote research and studies to increase the cyber range's effectiveness, such as examining the effectiveness of education and refining exercise scenarios based on behavioral analysis, to broaden the base of security personnel training.

\bibliographystyle{unsrt}  


\end{document}